\documentclass[prl,twocolumn,showpacs,superscriptaddress]{revtex4-1}

\usepackage{amsmath}
\usepackage[xdvi]{graphicx}%
\usepackage{dcolumn}
\usepackage{amsmath,amssymb}
\usepackage{color}
\usepackage{bm}

\usepackage{pdfpages}

\makeatletter
\AtBeginDocument{\let\LS@rot\@undefined}
\makeatother

\usepackage{color}
\definecolor{darkblue}{rgb}{0,0,0.6}
\definecolor{darkred}{rgb}{0.6,0,0}
\definecolor{darkgreen}{rgb}{0,0.6,0}

\usepackage[colorlinks=true,urlcolor=darkblue,citecolor=darkblue,linkcolor=darkred,hyperfootnotes=false]{hyperref}

\newcommand{\ee}{\mathrm{e}}

\begin{document}

\title{ 
Finite-Size Scaling of a First-Order Dynamical Phase Transition:\\
Adaptive Population Dynamics and an Effective Model
}


\author{Takahiro Nemoto}
\affiliation{Laboratoire de Probabilit\'es et Mod\`eles Al\'eatoires, Sorbonne Paris Cit\'e, UMR 7599 CNRS, Universit\'e Paris Diderot, 75013 Paris, France}
\affiliation{Philippe Meyer Institute for Theoretical Physics, Physics Department, \'Ecole Normale Sup\'erieure \& PSL Research University, 24 rue Lhomond, 75231 Paris Cedex 05, France}
\author{Robert L. Jack}
\affiliation{Department of Physics, University of Bath, Bath BA2 7AY, United Kingdom}
\author{Vivien Lecomte}
\affiliation{Laboratoire de Probabilit\'es et Mod\`eles Al\'eatoires, Sorbonne Paris Cit\'e, UMR 7599 CNRS, Universit\'e Paris Diderot, 75013 Paris, France}
\affiliation{LIPhy, Universit\'e Grenoble Alpes and CNRS, F-38042 Grenoble, France}

\date{\today}

\begin{abstract}
We analyze large deviations of the time-averaged activity in the one dimensional Fredrickson-Andersen model, both numerically and analytically.  The model exhibits a dynamical phase transition, which appears as a singularity in the large deviation function.  We analyze the finite-size scaling of this phase transition numerically, by generalizing an existing cloning algorithm to include a multi-canonical feedback control: this significantly improves the computational efficiency.  Motivated by these numerical results, we formulate an effective theory for the model in the vicinity of the phase transition, which accounts quantitatively for the observed behavior.  We discuss potential applications of the numerical method and the effective theory in a range of more general contexts.
\end{abstract}


\maketitle

{\it Introduction} --
Systems far from equilibrium display 
a wide spectrum of complex behavior~\cite{dorfman_introduction_1999,kampen_stochastic_2007}.  
For example, thermodynamic phase transitions are usually forbidden in one-dimensional systems, but a 
variety of \emph{dynamical phase transitions} are still observed~\cite{hinrichsen2000,evans2005,bertini_current_2005,bodineau_distribution_2005,garrahanjacklecomtepitardvanduijvendijkvanwijland}. 
Such transitions can appear in far-from-equilibrium states that are defined by restricting (or \emph{conditioning}) trajectories so that time-averaged observables take non-typical values~\cite{Touchette20091}.  They can be related to physical properties of systems where metastability is important, especially glassy systems~\cite{merolle_spacetime_2005,jack_space-time_2006,garrahanjacklecomtepitardvanduijvendijkvanwijland,garrahan_first-order_2009,kurchan_six_2009,hedges_dynamic_2009,garrahan_kinetically_2010,chandler_dynamics_2010,limmer2014}.  

In some cases these transitions can be studied analytically~\cite{bodineau_cumulants_2007,bertini_current_2005,bertini_non_2006,bodineau_distribution_2005,garrahanjacklecomtepitardvanduijvendijkvanwijland}, but in practical applications one must often resort to numerical methods: These access the relevant far-from-equilibrium states by \emph{rare-event sampling} algorithms~\cite{giardina_direct_2006,hedges_dynamic_2009,pitard_dynamic_2011,speck2012,malins2012}, employing for instance population dynamics or path sampling.
Such methods tend to perform poorly in the vicinity of dynamical phase transitions, just as conventional sampling methods tend to fail close to equilibrium phase transitions.  For the equilibrium case, advanced methods exist that solve this problem, including finite-size scaling analysis~\cite{privmanfisher,borgskotecky} and multi-canonical sampling~\cite{PhysRevLett.68.9,PhysRevLett.86.2050,bruce2003}.  For dynamical phase transitions, some progress has been made in this direction~\cite{speck2012,limmer2014,gingrich2015}  but accurate calculations are numerically expensive and suffer from significant finite-size effects.

Here, we analyze a dynamical phase transition~\cite{garrahanjacklecomtepitardvanduijvendijkvanwijland} in the Fredrickson--Andersen (FA) model~\cite{fredrickson_kinetic_1984}. We combine a state-of-the-art numerical approach~\cite{NemotoBouchetLecomteJack} with a theoretical analysis. 
 We show that numerical results and theoretical predictions for finite-size scaling near the phase transition agree quantitatively. By combining these ingredients we obtain a full description of the transition, at a modest computational cost.  The phase transition is a prototype for transitions in a range of systems~\cite{garrahanjacklecomtepitardvanduijvendijkvanwijland,elmatad2010,hedges_dynamic_2009,limmer2014}, so we argue that these new methods and insights have broad potential application in this field.

{\it Model} --
The one-dimensional (1d) FA model~\cite{fredrickson_kinetic_1984} is a kinetically constrained model (KCM) that 
consists of $L$ spins on a periodic lattice.  The $i$th spin takes values $n_i=0$ (down) or $n_i=1$ (up) and the configuration is $\mathcal{C}=(n_i)_{i=1}^{L}$.  We define an  
 operator $\mathcal F_i$ that flips the state of spin $i$, so that $\mathcal F_i [ \mathcal C] = (n_1,n_2,\cdots,1-n_i,\cdots,n_L)$.  The kinetic constraint of the model is that spin $i$ can flip only if at least one of its neighbors is up.  The transition rates between configurations reflect this constraint, they are:
\begin{equation}
w(\mathcal C \rightarrow \mathcal F_i [\mathcal C] ) =\left [ c \left (1 - n_i \right ) + \left (1-c \right ) n_i  \right ] f_i(\mathcal C),
\end{equation}
where $f_i=n_{i-1}+n_{i+1}$ enforces the kinetic constraint and $c$ is a parameter that depends on the temperature in the model~\cite{fredrickson_kinetic_1984}.
The rates obey detailed balance and the model's equilibrium distribution follows a Bernoulli law, $p_{\rm eq}(\mathcal C) \propto c^{\sum_i n_i} (1-c)^{L-\sum_i n_i}$. %
Despite this trivial distribution, the kinetic constraint in the model leads to rich behavior, related to  dynamical heterogeneity in glassy systems~\cite{doi:10.1080/0001873031000093582,garrahan_kinetically_2010,hedges_dynamic_2009}. 

\emph{Dynamical phase transitions} --
We define the \emph{dynamical activity} $K(\tau)=N^{\rm K}(\tau)/\tau$ where $N^{\rm K}(\tau)$ is the total number of spin flips during the time interval $[0,\tau]$.  The phase transitions that we consider take place in ensembles of trajectories that are restricted to a given value of the activity.  In the limit $\tau\to\infty$, the activity converges to its equilibrium value $K_{\rm eq}$: to estimate the probability of rare trajectories with $K(\tau)\neq K_{\rm eq}$, we consider the cumulant generating function (CGF)
\begin{equation}
G ( s ) =  \lim _{\tau \rightarrow \infty} \frac{1}{\tau } \log \left \langle {\rm e}^{-s \tau K(\tau)} \right \rangle,
\label{eq:defGen}
\end{equation}
where $\left \langle \cdot  \right \rangle$ denotes an ensemble average.  Dynamical phase transitions are associated with singularities in $G(s)$: they are analogous to thermodynamic phase transitions, with $G$ corresponding to the thermodynamic free energy~\cite{garrahan_first-order_2009} and $s$ corresponding to an intensive thermodynamic field that is used to drive the system through its phase transition.
We also define
\begin{equation}
\left \langle K \right \rangle _s  \equiv  -\frac{\mathrm{d}G(s)}{\mathrm{d}s} =  \lim_{\tau \rightarrow \infty} \frac{ \left \langle K(\tau){\rm e}^{-s \tau K(\tau)}  \right \rangle }{\left \langle {\rm e}^{-s \tau K(\tau)}  \right \rangle},
\label{eq:defK}
\end{equation}
which specifies the dependence of the mean activity on the field $s$, analogous to the dependence of the order parameter on its conjugate field in thermodynamics.

\begin{figure}
\centering
\includegraphics[width=8.4cm]{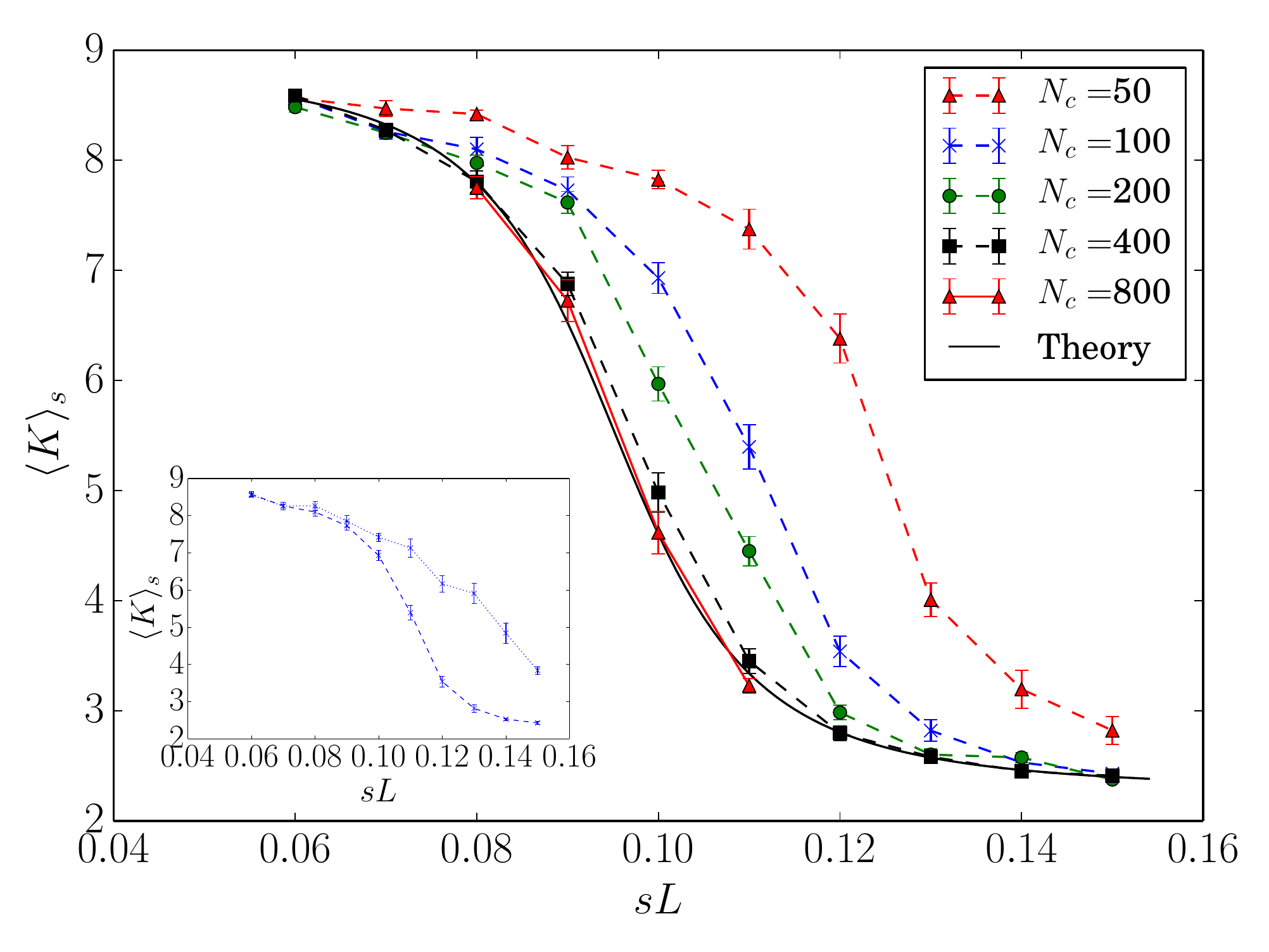}
\caption{\label{Fig:ScalingFunction}%
The average activity $\left \langle K \right \rangle_s$ for $L=36$ as a function of $sL$, estimated using the feedback method described in the text. 
As the number of copies $N_c$ increases, the estimators of $\left \langle K \right \rangle_s$ converge to the correct result. 
The solid black line is  the analytical form (\ref{scalingfunctionanaly}). The parameters $A, B, \kappa, s_c^{L}$ in (\ref{scalingfunctionanaly}) are determined by fitting the data {\it outside} of the coexistence region ($sL=0.06\sim 0.08$, $0.12 \sim 0.15$) for $N_c=400$, where the method converges rapidly. The inset shows a comparison between the results obtained from the feedback method (blue dashed line) and from the standard method (blue dotted line) for $N_c=100$. 
}
\end{figure}

The dynamical phase transition that occurs in the FA model separates a high-activity state [with $\langle K\rangle_s=O(L)$] from an inactive (glass) state [with $\langle K\rangle_s=o(L)]$. It is defined in a joint limit of large time $\tau$ and large system size $L$.
In this work, we first take $\tau\to\infty$ and then take $L\to\infty$.
The phase transition is first-order, and the order parameter $K^\infty(s)=\lim_{L\to\infty} \frac 1L \left \langle K \right \rangle _s$ exhibits a discontinuous jump at $s=0$~\cite{garrahanjacklecomtepitardvanduijvendijkvanwijland}.  However, the large-$L$ limit is not accessible numerically, and for finite $L$ the activity is a smooth function of $s$, whose representative behavior is shown in Fig.~\ref{Fig:ScalingFunction}.  The crossover sharpens as $L$ increases: to analyze the transition, one must consider the finite-size scaling of $\langle K \rangle_s$.

{\it Cloning algorithm with feedback} -- 
To perform this finite-size scaling, we require a numerical method that provides accurate results for a range of system sizes.  To this end, we generalize a recently-proposed \emph{adaptive method}~\cite{NemotoBouchetLecomteJack} to Markov jump processes.  The method is based on a
\emph{cloning} algorithm~\cite{giardina_direct_2006,giardina_simulating_2011} which uses a population of $N_c$ clones (or copies) of the system.  We fix a time interval $\Delta t$ and the dynamics of the model are propagated over intervals of length $\Delta t$, such that the total time is $\tau$.  For each interval, one calculates a weighting factor for clone $a$:
\begin{equation}
\gamma_a = \exp \left \{  -s \left [ N^{\rm K}_a(t+\Delta t) - N^{\rm K}_a(t) \right ]  \right \},
\label{eq:originalweightingratio}
\end{equation}
where $ N^{\rm K}_a(t)$ is the number of spin flips for clone $a$, evaluated over the whole time interval $[0,t]$; also one defines $K_a(t)\equiv N^{\rm K}_a(t)/t$.
After each time interval, clones are duplicated or removed, to enforce the conditioning on the activity.  In this step, each clone $a$ generates a number of offspring proportional to its weight $\gamma_a$.
The mean activity $\langle K \rangle_s$ can then be obtained as the average of $K_a(\tau)$ over the final population~\cite{NemotoBouchetLecomteJack}.

This algorithm provides accurate results when $N_c$ is sufficiently large~\cite{NemotoGuevaraLecomte}, but in practice this may require a very large number of clones, which is computationally expensive. 
To avoid this issue, we combine the existing cloning algorithm~\cite{giardina_direct_2006,giardina_simulating_2011,hedges_dynamic_2009,pitard_dynamic_2011,speck2012,malins2012,NemotoGuevaraLecomte} with 
a modification of the dynamics \cite{jack_large_2010,1742-5468-2010-10-P10007,PhysRevLett.111.120601,PhysRevLett.112.090602}, following~\cite{NemotoBouchetLecomteJack}. In order to aid sampling of trajectories with non-typical activity, we modify the transition rates of the model as
\begin{equation}
w_{\rm mod} (\mathcal C \rightarrow \mathcal F_i[ \mathcal C])   =  {\rm e}^{-s} w(\mathcal C \rightarrow \mathcal F_i[ \mathcal C]) {\rm e}^{\frac12[U({\cal C}) - U({\cal F}_i [{\cal C}])]},
\label{eq:wmod}
\end{equation}
where $U({\cal C})$ is an {effective potential} or \emph{control potential}~\cite{jack2015_eff}.   The weight factors $\gamma_a$ are also modified, by replacing $-s N^{\rm K}$ in (\ref{eq:originalweightingratio}) with
\begin{equation}
K_{\rm mod} = \int_{0}^{\tau} dt   \left [  k_{\rm mod}(\mathcal C_t) - k(\mathcal C_t)   \right ],
\label{eq:Modifiedbias}
\end{equation}
where $k(\mathcal C)=\sum_i w (\mathcal C \rightarrow \mathcal F_i[ \mathcal C])$ is the escape rate from configuration $\cal C$, and $k_{\rm mod}$ is obtained in the same way but using the modified rates~\eqref{eq:wmod}.  

In the limit of large  $N_c$, the results of the algorithm are independent of the choice of $U$.  However, an appropriate choice can dramatically improve the accuracy of results obtained with finite populations.  
In particular, there exists an optimal control potential for which a population of $N_c=1$ is already sufficient for convergence.  
This optimal potential is given (up to an arbitrary constant) by $U^*({\cal C})=2 \log[ p_{\rm eq}(\mathcal C)/p_{\rm end}({\cal C})]$ where $p_{\rm end}(\cal{C})$ is the probability of observing configuration $\cal C$ within the steady state of the cloning algorithm~\cite{NemotoBouchetLecomteJack}.  Calculating this optimal control directly is not feasible in practice: instead we restrict to control potentials that involve interactions between each spin and its nearest neighbors at distance~$\leq d$, so that the  change in the effective potential on flipping spin $i$ is
\begin{equation}
U(\mathcal F_i[\mathcal C]) - U(\mathcal C) = u_d(n_{i-d},\dots,n_i,\dots,n_{i+d})
\label{equ:dU}
\end{equation}
for some function $u_d$.
To obtain the most suitable values for these potentials, we use a \emph{feedback scheme}: we run the cloning algorithm, estimate the probabilities of particular local arrangements of the spins, and update the effective potential based on that choice.  By repeating this procedure, one can optimize the choice of the control potential (see~\cite{NemotoBouchetLecomteJack} and~\cite{SM1} for details).

{\it Results} -- 
Fig.~\ref{Fig:ScalingFunction} shows the performance of the algorithm, close to the dynamical phase transition. We plot $\left \langle K \right \rangle_s$ as a function of $s L$, 
as obtained from the cloning algorithm, using the feedback method to determine suitable effective interactions.  The interaction range is $d=4$, and we take $N_c$ between 50 and 800.  We set $\tau = 15000$, which is sufficiently large to converge to the large-$\tau$ limit.  As $N_c$ increases, the estimates of $\langle K \rangle_s$ converge to a smooth curve, indicating that these clone populations are large enough to achieve accurate results.  By contrast, the inset to Fig.~\ref{Fig:ScalingFunction} shows that the original cloning method (with $U=0$) deviates significantly from the correct result, compared with the feedback method for the same number of copies. 
This tendency is observed throughout the whole range of $s$, irrespective of the presence of the dynamical coexistence~\cite{SM2}.

\begin{figure}
\centering
\includegraphics[width=8cm]{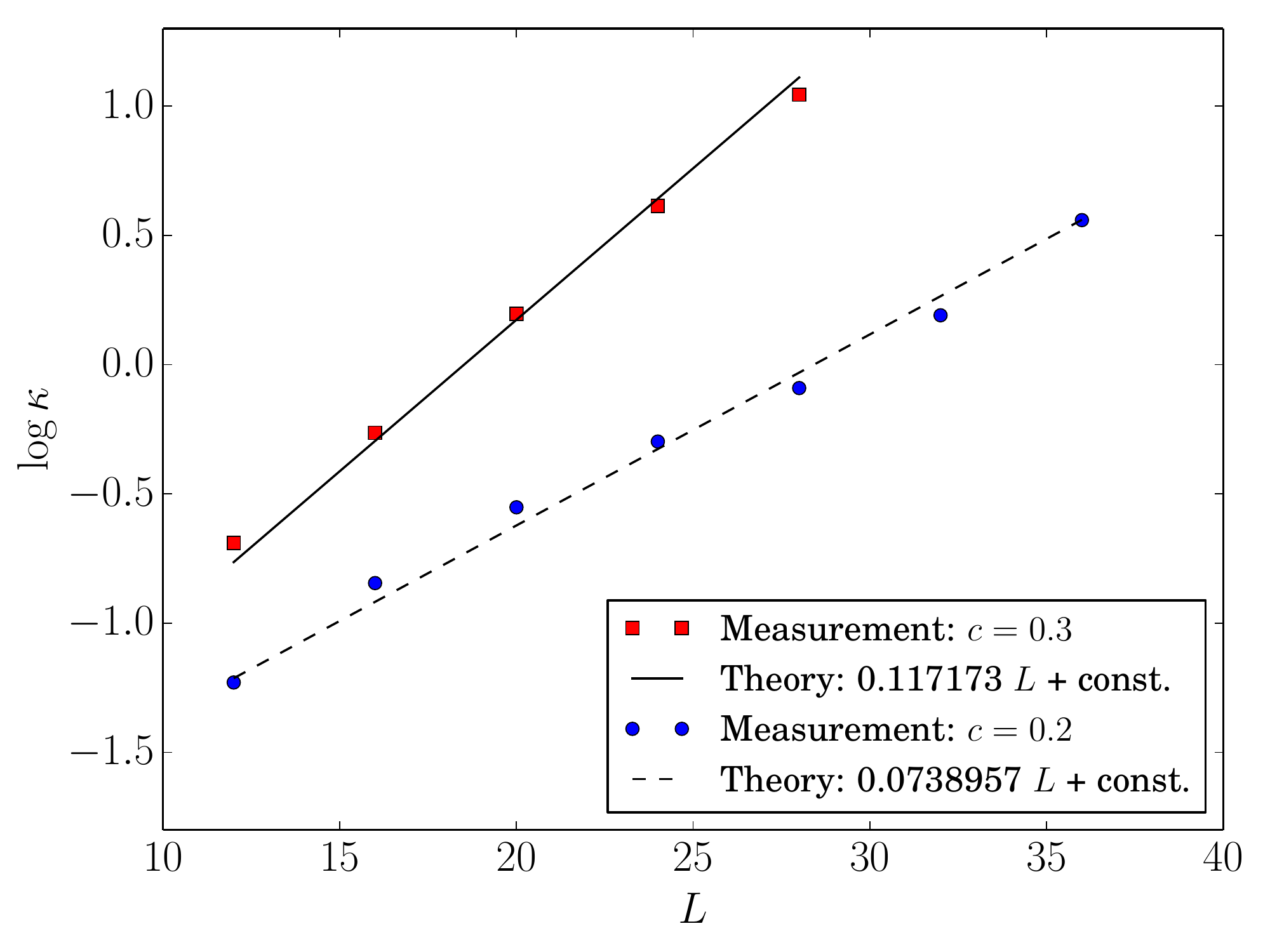}
\caption{\label{Fig:Scalingspeed} Exponential divergence of the dynamical susceptibility $\kappa$, defined in (\ref{def:eqKappa}). We plot $\log \kappa$  as a function of $L$, as obtained from the feedback method (points) together with the theoretical prediction (\ref{eq:exponentformula}) (straight lines). 
}
\end{figure}

The results of Fig.~\ref{Fig:ScalingFunction} show the expected crossover from high to low activity, consistent with the existence of a dynamical phase transition near $s=0$.  To analyze the finite-size scaling of this transition, we define $\kappa$ as the maximal susceptibility
\begin{equation}
\kappa \equiv  \frac{1}{L^2}\max_{s } \left| \frac{\partial \left \langle K \right \rangle_s}{\partial s} \right|.
\label{def:eqKappa}
\end{equation}
Let the finite-size transition point $s_c^L$ be the value of $s$ at which this maximum occurs. For large systems, we expect $\kappa\to\infty$ and $s_c^L =\mathcal O(L^{-1})$~\cite{garrahanjacklecomtepitardvanduijvendijkvanwijland,bodineautoninelli}.
Fig.~\ref{Fig:Scalingspeed} shows the dependence of $\kappa$ on the system size $L$: the results are consistent with an exponential divergence of $\kappa$ as $L \rightarrow \infty$, in contrast to traditional finite-size scaling at thermodynamic transitions, where $\kappa$ scales as a power of $L$~\cite{borgskotecky,elmatad2010}.  To gain insight into these phase transitions and explain the numerical results in Figs.~\ref{Fig:ScalingFunction} and~\ref{Fig:Scalingspeed}, we now present some theoretical arguments.

\begin{figure}
\centering
\includegraphics[width=8.3cm]{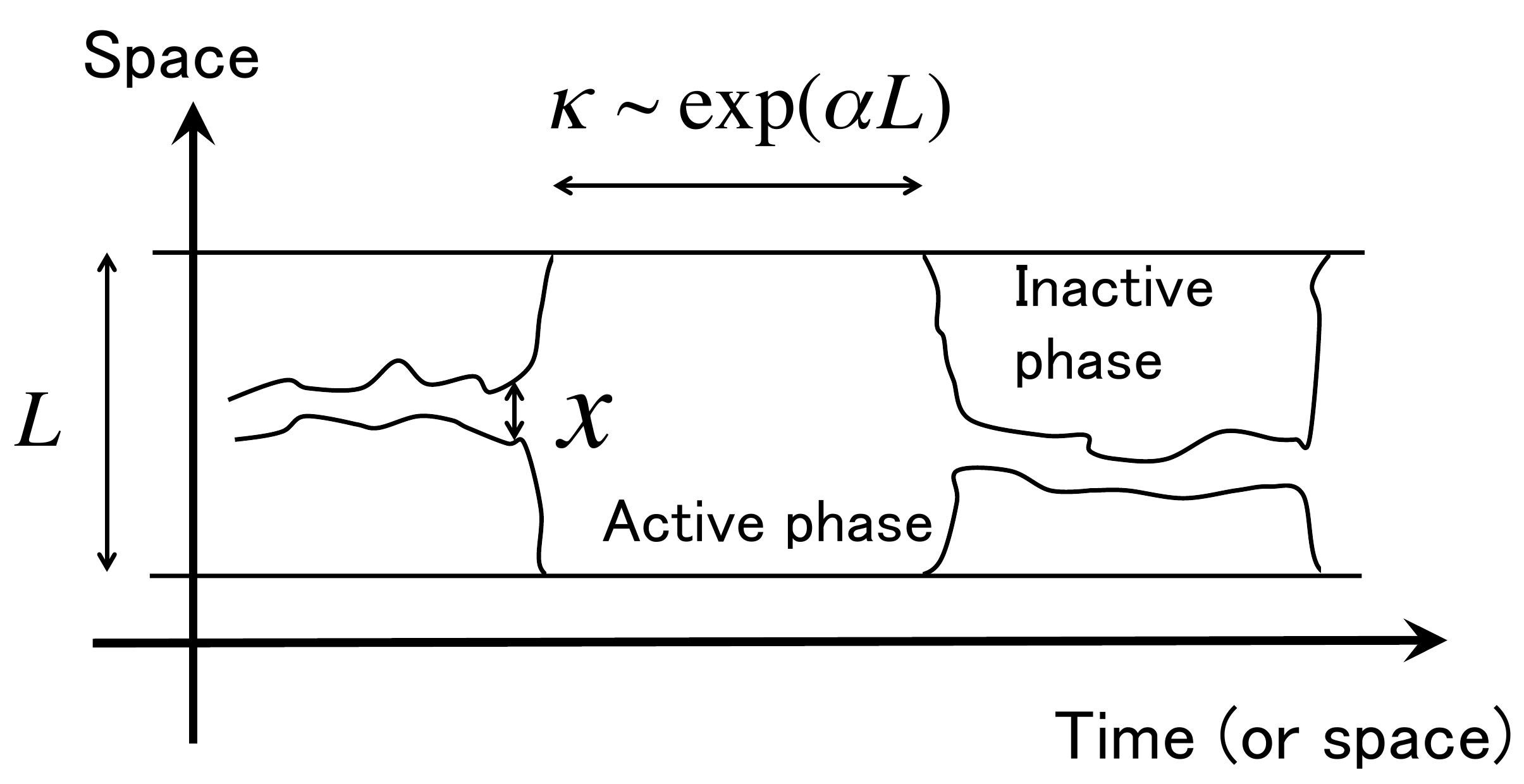}
\caption{\label{Fig:schematic} Schematic picture of the domain wall dynamics at dynamical phase coexistence, which is analogous to  equilibrium phase coexistence on the (2d) surface of a long cylinder~\cite{privman_finite-size_1983,borgs_crossover_1992} (in this latter case the horizontal axis is a spatial co-ordinate). 
In the effective model, the fluctuating variable $x \in \{1,2,\cdots,L\}$  represents the width of the active phase.  
}
\end{figure}

{\it Analogy with a two-dimensional thermodynamic system on a cylinder} --  These dynamical phase transitions in one dimension can be mapped to thermodynamic transitions in two dimensions (2d)~\cite{jack_large_2010,elmatad2010}.  Recalling that we have taken the limit $\tau\to\infty$ before taking $L\to\infty$, the relevant geometry for the 2d system is a long cylinder, with the system size $L$ in the dynamical system corresponding to the perimeter of the cylinder.  For equilibrium systems in such geometries, the behavior near phase coexistence is sketched in Fig.~\ref{Fig:schematic}~\cite{privman_finite-size_1983,borgs_crossover_1992}: the two phases form domains arranged along the cylinder.  The typical domain length scales exponentially in $L$: the reason is that these 2d domains are separated by domain walls of length $L$ which run around the cylinder, and the associated interfacial free-energy cost scales as $\alpha L$, so the density of domain walls is of order $\ee^{-\alpha L}$.   From (\ref{def:eqKappa}), $\kappa$ is analogous to a susceptibility in the thermodynamic transition; it is also equal to a time-integral of the autocorrelation function of $k(\mathcal C_t)/L$~\cite{garrahan_first-order_2009}.  Hence $\kappa$ scales with the relaxation time of the system.
Identifying this relaxation time with the domain size in Fig.~\ref{Fig:schematic}, the susceptibility therefore diverges 
as $\kappa=O({\rm e}^{\alpha L})$ for large $L$.  The numerical data in Fig.~\ref{Fig:Scalingspeed} are consistent with such a divergence, indicating that the thermodynamic analogy can predict properties of the dynamical transition. 

{\it Effective interfacial model of the dynamical phase transition} --
We now introduce a simplified effective model for the domains in Fig.~\ref{Fig:Scalingspeed}.  
Following Section~4.1 of~\cite{bodineaulecomtetoninelli}, we assume that typical configurations in the model include a single active domain of size $x$ (Fig.~\ref{Fig:schematic}).  Within this domain, the system is close to its active (equilibrium) state; in the remainder of the system, the system is inactive and there are no up spins.  The system contains at least one up spin so $1\leq x\leq L$.  We will show that this simplified model makes quantitatively accurate predictions for the dynamical phase transition.  (For thermodynamic transitions, similar results may be available via spectral properties of the transfer matrix~\cite{privman_finite-size_1983,borgs_crossover_1992}: our analysis here is different, and is based on the dynamical nature of the phase transition.)

The dynamical rules of the FA model mean that the value of $x$ increases with rate $2c(1-c)$ and decreases with rate $2c$~\cite{bodineaulecomtetoninelli}. We take reflecting boundary conditions at $x=1,L$.  (For systems that are predominately active, we interpret $(L-x)$ as the size of the largest inactive domain in the system, which has a typical value of order $1/c$.)  The activity in this effective model is obtained by assuming that the spins in the active domain flip with typical rate $\overline k = 4 c^2(1-c)$, but there are no spin flips outside this region, due to the kinetic constraint.  Hence the analog of $K(\tau)$ is $(\overline k/\tau) \int_0^\tau  x(t) dt$.

The result is an effective model where $x$ undergoes a random walk whose hop rates are biased to the right (positive $x$), but conditioned on a relatively small time-averaged position.
In the limit of large $L$, the model can be solved exactly, as shown in supplemental material~\cite{SM3}. We summarize the main results: There is a dynamical phase transition at a field $s^* = \mu^3/(L\sqrt{\overline k})$, where $\mu$ solves
\begin{equation}
\frac12 {\mu^{3/2}} + F(c) =   \frac{ 4c^2  - \overline k + 2c \sqrt{\overline k}  \Psi(c,\mu)  }{ 4 c^2 + \overline k - 2c\sqrt{\overline k}\Psi(c,\mu) }.  
\label{eq:analycicalkappa}
\end{equation}
Here, $F(c) =(1/2) \log\left [ c/(1-c) \right ]$ and $\Psi(c,\mu)=   -  \mu^3  + 2( \cosh F(c) -1)$. 
Also, the susceptibility $\kappa$ diverges as 
\begin{equation}
\kappa \propto {\rm e}^{\alpha L}, \qquad \alpha=\tfrac23\mu^{3/2}.
\label{eq:exponentformula}
\end{equation} 
Finally, the scaling form of the activity near the phase transition is
 \begin{equation}
\frac{\left \langle K \right \rangle _s}{L}  \sim A  - \frac{\kappa L (s- s_c^L) }{ \sqrt{1+ B \kappa^2 L^2 ( s-  s_c^L)^2 }},
\label{scalingfunctionanaly}
\end{equation}
where $A$ and $B$ are specified in~\cite{SM3}. 
This last result is similar to that obtained in a mean-field FA model~\cite{1742-5468-2014-10-P10001} and that for 2d equilibrium phase coexistence of ferromagnets on a cylinder geometry~\cite{borgs_crossover_1992}. 
These similarities indicate that the scaling function (\ref{scalingfunctionanaly}) might be a general property of first-order phase transitions with exponentially diverging susceptibilities.  We also remark that our effective model yields the interfacial free-energy cost $\alpha$ (\ref{eq:analycicalkappa},\ref{eq:exponentformula}), which is not available from the 
2d equilibrium approach of Refs.~\cite{privman_finite-size_1983,borgs_crossover_1992}.

The theoretical predictions (\ref{eq:analycicalkappa}-\ref{scalingfunctionanaly}) are shown in Figs.~\ref{Fig:ScalingFunction} and~\ref{Fig:Scalingspeed}, together with the numerical results.  The agreement is excellent, despite the simplicity of the model.  The conclusion is that the finite-size scaling of the phase transition is dominated by the dynamical properties of the interface between the active and inactive regions, and this interface is accurately described by the effective model.  Moreover, in the analogy with the classical phase transition on a cylinder, we can interpret the parameter $\alpha$ in terms of an interfacial tension between the active and inactive domains shown in Fig.~\ref{Fig:Scalingspeed}.

\emph{Discussion} -- There are two key outcomes of this work.  First, we have shown that the cloning-with-feedback algorithm used here allows accurate characterization of dynamical phase transitions for a range of system sizes, with much greater computational efficiency than the original cloning scheme.  Second, we have shown how the finite-size scaling of first-order dynamical phase transitions can be understood qualitatively by mapping them to classical phase transitions in cylindrical geometries; it can also be analyzed \emph{quantitatively} by mapping to the effective interfacial model.  

We expect both the numerical and theoretical methods to apply generally for dynamical phase transitions of this type: for example, application to other KCMs~\cite{garrahanjacklecomtepitardvanduijvendijkvanwijland,elmatad2010} 
should be straightforward.  We also anticipate application to atomistic systems that support similar phase transitions~\cite{hedges_dynamic_2009,pitard_dynamic_2011,speck2012,malins2012,limmer2014}. Moreover, the transitions considered here are directly related to quantum phase transitions in spin chains, for which results similar to (\ref{eq:analycicalkappa}) have been derived~\cite{heyl_dynamical_2013}.  The effective interfacial model presented here provides a clear physical interpretation of such results, whose implications for quantum systems remain to be explored.

We also highlight several useful features of the numerical algorithm used here.  The computational cost of the cloning algorithm scales linearly in the time $\tau$.  This allows the large-$\tau$ limit to be converged numerically.  Hence, the only parameter in the finite size scaling is $L$, which allows direct comparison with the theory presented here.  This analysis is significantly simpler than finite-size scaling via path sampling, where both $\tau$ and $L$ must be varied together~\cite{hedges_dynamic_2009,speck2012}.  The cloning algorithm can also be applied in systems where detailed balance is broken, where path-sampling methods are not directly applicable.  Cloning methods are also related to Diffusion Quantum Monte Carlo~\cite{Anderson1975_63_4}: it would be interesting to investigate how the cloning-with-feedback method might be applied in that context~\cite{casula_beyond_2006}.

Another advantage of this method is that the control potential $U$ determined numerically provides physical insight into these dynamical phase transitions.  In the active phase close to the transition, the control potential acts to suppress the number of up spins (reducing the activity), but one also finds an effective attraction between up spins~\cite{SM4}.  This attraction is weak but decays slowly in space, which acts to stabilize the large spatial domains shown in Fig.~\ref{Fig:schematic}~\cite{jack2014_east,jack2015_hyper}.
Based on the effective model, we
expect that the optimal control potential $U^*$ should depend primarily on these domain sizes, so it naturally includes long-range interactions.
It would be interesting to obtain a better understanding of optimal control potentials close to dynamical phase transitions, especially since incorporating such information into numerical methods now has the potential to significantly improve their performance.  For example, one might consider transitions in other spin models~\cite{garrahan_first-order_2009} as well as exclusion processes~\cite{bodineau_distribution_2005,jack2015_hyper,1751-8121-45-17-175001}.
In any case, we stress that while the ansatz~(\ref{equ:dU}) is much simpler than the optimal control, it still results in a significant improvement of the computational efficiency.

In conclusion, we have shown how a combination of numerical and theoretical methods provide a detailed insight into the dynamical phase transition in the FA model.  
The basic ideas of the method are quite general, such as the modification of the cloning algorithm with a feedback procedure to determine the optimal force, or the interfacial model as a coarse-grained description of systems near coexistence.
For first-order dynamical transitions, we believe that effective interfacial models should apply rather generally.
The numerical method has even broader potential application, although the choice of a suitable control potential will depend on the problem of interest -- this remains to be explored.
%
%
%

\begin{acknowledgments}
T.~N.~gratefully acknowledges the  support of Fondation Sciences Math\'ematiques de Paris -- EOTP NEMOT15RPO, PEPS LABS and LAABS Inphyniti CNRS project.
V.~L.~acknowledges support by the ANR-15-CE40-0020-03 Grant LSD and by the ERC Starting Grant 680275 MALIG.
T.~N.~and V.~L.~are grateful to G.~Semerjian for discussions.

\end{acknowledgments}

\bibliographystyle{plain_url}
\bibliography{draft.bib}



\section*{Supplementary material}

# Supplementary materials for "Finite-size scaling of a first-order dynamical phase transition: adaptive population dynamics and effective model"

Takahiro Nemoto, Robert L. Jack, and Vivien Lecomte

This supplementary material is constituted of four parts. In the first part, we explain the details of the feedback algorithm used in the main text. In the second part, we show numerical evidence supporting the feedback algorithm compared to the standard one, in the active phase. In the third part, we discuss the random-walk effective model to describe the dynamics of the interface separating the active and the inactive domains, in the phase-coexistence regime. We derive the relations (9), (10) and (11) of the main text, which characterize the exponential scaling around the phase transition. Finally, we study the effective interactions in the vicinity of the dynamical phase transition and describe the corresponding underlying physical picture.

**Contents**

# I. CLONING ALGORITHM WITH A FEEDBACK PROCEDURE

In this first part, we describe in detail the feedback algorithm. In the first subsection, we summarize the theoretical background at the basis of the algorithm [1], and in the next subsection, we provide its implementation. It alternates iteratively (i) the estimation of a probability distribution and (ii) an update of the transition rates, based on this estimation.

## A. Analytical background

### 1. Equivalence between two population dynamics

As described in the main text (around Eq. (5)), we use a dynamics whose transition rates are modified by a potential $U(\mathcal{C})$. Following Ref. [1], we consider the ratio of path probabilities for the original dynamics (denoted by $P[(\mathcal{C}_t)_{t=0}^\tau]$) and for the modified dynamics (denoted by $P_U[(\mathcal{C}_t)_{t=0}^\tau]$). The ratio is calculated from the relation

$$P[(\mathcal{C}_t)_{t=0}^\tau]e^{-s\tau K(\tau)} = e^{(1/2)[U(\mathcal{C}(\tau))-U(\mathcal{C}(0))]} P_U[(\mathcal{C}_t)_{t=0}^\tau]e^{K_{\mathrm{mod}}(\tau)}, \tag{S1}$$

where $K_{\mathrm{mod}}(\tau)$ is a modified bias, given by Eq. (6) of the main text. The interpretation of this equation is simple. In the large observation time limit $\tau \to \infty$, the potential difference $U(\mathcal{C}(\tau)) - U(\mathcal{C}(0))$ between the initial and the final times can be neglected. The following two population dynamics thus become equivalent: (i) the original population dynamics biased with $-s\tau K(\tau)$ and (ii) the population dynamics with rates modified by $U(\mathcal{C})$, now biased with $K_{\mathrm{mod}}(\tau)$. Our aim is to estimate numerically the "best" potential $U(\mathcal{C})$ which renders the modified dynamics more efficient, in a sense that is specified below.

### 2. Two distribution functions defined in population dynamics algorithm

In the population dynamics algorithm, we define the following two distributions [1]: the final-time distribution

$$p_{\mathrm{end}}(\mathcal{C}) = \lim_{\tau, N_c \to \infty} \frac{1}{\tau N_c} \sum_{a=1}^{N_c} \int_0^\tau dt\, \delta_{\mathcal{C},\mathcal{C}_t^a}, \tag{S2}$$

where $\mathcal{C}_t^a$ is the trajectory of $a$th clone generated from the population dynamics, and the intermediate-time distribution

$$p_{\mathrm{ave}}(\mathcal{C}) = \lim_{\tau, N_c \to \infty} \frac{1}{\tau N_c} \sum_{a=1}^{N_c} \int_0^\tau dt\, \delta_{\mathcal{C},\tilde{\mathcal{C}}_{t,\tau}^a}, \tag{S3}$$

where $\tilde{\mathcal{C}}_{t,\tau}^a$ is the trajectory of $a$th clone which has survived until the final time $\tau$. One can show that an *optimal potential* $U^*(\mathcal{C})$ defined from the ratio of these two distributions as [1]

$$U^*(\mathcal{C}) = -2\log \frac{p_{\mathrm{ave}}(\mathcal{C})}{p_{\mathrm{end}}(\mathcal{C})} \tag{S4}$$

possesses a remarkable property: In the dynamics modified with $U^*(\mathcal{C})$, the bias $K_{\mathrm{mod}}(\tau)$ is a constant (proportional to $\tau$) that does not depend on the trajectory $(\mathcal{C}_t)_{t=0}^\tau$. This means that the population dynamics modified by $U^*$ presents a uniform cloning, implying that it can be simulated without resorting to any population dynamics. This property would render the cloning algorithm trivial, but would rely on an *exact* evaluation of $U^*(\mathcal{C})$, which is a difficult task, both numerically and analytically. Indeed, one can show that the optimal potential is directly expressed in terms of the eigenvector corresponding to the CGF, seen as an eigenvalue of the biased evolution operator. To proceed, the idea is to evaluate an approximation of $U^*(\mathcal{C})$, that will still render the cloning algorithm more efficient.

For this purpose, we now relate $U^*(\mathcal{C})$ to the final- and intermediate-time distributions in the dynamics modified by an arbitrary potential $U(\mathcal{C})$, which we denote respectively by $p_{\mathrm{end}}^U$ and $p_{\mathrm{ave}}^U$. They are related to the original ones as $p_{\mathrm{end}}^U = p_{\mathrm{end}} e^{-U/2}$ and $p_{\mathrm{ave}}^U = p_{\mathrm{ave}}$ [1], which means that (S4) can be rewritten in terms of $p_{\mathrm{end}}^U$ and $p_{\mathrm{ave}}^U$ as:

$$U^*(\mathcal{C}) = U(\mathcal{C}) - 2\log \frac{p_{\mathrm{ave}}^U(\mathcal{C})}{p_{\mathrm{end}}^U(\mathcal{C})}. \tag{S5}$$

We note that in certain conditions, one can express $p_{\text{ave}}$ by using $p_{\text{end}}$. For example, when the modified transition rates $w_{\text{mod}}(\mathcal{C} \to \mathcal{C}')$ satisfy the detailed balance condition, then $w_{\text{mod}}(\mathcal{C} \to \mathcal{C}')$ satisfy the relation $p_{\text{ave}}(\mathcal{C})w_{\text{mod}}(\mathcal{C} \to \mathcal{C}') = p_{\text{ave}}(\mathcal{C}')w_{\text{mod}}(\mathcal{C}' \to \mathcal{C})$. By combining this detailed balance condition with the one of the original system, namely $p_{\text{eq}}(\mathcal{C})w(\mathcal{C} \to \mathcal{C}') = w(\mathcal{C}' \to \mathcal{C})p_{\text{eq}}(\mathcal{C}')$, we obtain $p_{\text{ave}}(\mathcal{C}) = p_{\text{eq}}(\mathcal{C})e^{-U(\mathcal{C})}$. This leads to $p_{\text{ave}}(\mathcal{C}) = p_{\text{end}}(\mathcal{C})^2/p_{\text{eq}}(\mathcal{C})$, or equivalently,

$$p_{\text{ave}}^U(\mathcal{C}) = \frac{p_{\text{end}}^U(\mathcal{C})^2}{p_{\text{eq}}(\mathcal{C})} e^{U(\mathcal{C})} . \tag{S6}$$

When detailed balance holds, this allows us to express the optimal potential through (S5) in terms of the final-time distribution $p_{\text{end}}^U$ only. This simplifies the numerical procedure, by avoiding the estimation of $p_{\text{ave}}^U$.

### B. Algorithm

The basic idea of the feedback algorithm is to estimate iteratively an approximation $U(\mathcal{C})$ of the optimal potential $U^*(\mathcal{C})$, using, based on (S5), the numerical measurement of a certain probability distribution along the population dynamics. Specifically, we look for a modifying potential $U$ such that

$$U(\mathcal{F}_i[\mathcal{C}]) - U(\mathcal{C}) = u_d(n_{i-d}, \cdots, n_i, \cdots, n_{i+d}). \tag{S7}$$

where $\mathcal{C}$ denotes the configuration of the $n_j$'s and $\mathcal{F}_i$ is the operator which flips a spin at site $i$. Such form represents interactions in the modified dynamics occurring only at a bounded range $d$. In order to evaluate $u_d$, we define two empirical distribution functions for blocks of spins of length $(2d+1)$:

$$p_{\text{end}}^d(\mathcal{C};i) = \frac{1}{\tau N_c} \sum_{a=1}^{N_c} \int_0^\tau dt \prod_{j=i-d}^{i+d} \delta_{n_j,(n_j(t))_a} , \tag{S8}$$

where $(n_i(t))_a$ is the configuration of $i$th spin at time $t$ for the clone $a$, and

$$p_{\text{ave}}^d(\mathcal{C};i) = \frac{1}{\tau N_c} \sum_{a=1}^{N_c} \int_0^\tau dt \prod_{j=i-d}^{i+d} \delta_{n_j,(\tilde{n}_j(t,\tau))_a} , \tag{S9}$$

where $(\tilde{n}_i(t,\tau))_a$ is the configuration of $i$th spin at time $t$ for the $a$th clone *surviving* up to the final time $\tau$ [1]. We note that we measure $p_{\text{end,ave}}^d$ on a fixed site $i$, and use the obtained result for every other site $1, 2, \cdots, L$, assuming a translational invariance. But in general, due to numerical errors, $p_{\text{end,ave}}^d(C;i)$ can take different values depending on $i$. Choosing a long enough duration for the measurement of $p_{\text{end,ave}}^d(C;i)$, these numerical fluctuations are negligible in our case. However, in other cases where the duration for the measurement of $p_{\text{end}}^d(C;i)$ is too short, one should measure $p_{\text{end,ave}}^d(C;i)$ for all $i$ and use the corresponding translationally averaged value.

We iterate the following procedure until the results of the algorithm converge. We denote the index of the iterations by $\ell$. The variation (S7) of effective potential is updated at the end of each iteration, and is denoted by $u_d^\ell$. Initially, $u_d^{\ell=0} = 0$. We iterate the following steps:

1. We perform the population dynamics algorithm with the modifying potential $u_d^\ell$ and measure $p_{\text{end}}^d$ and $p_{\text{ave}}^d$ according to (S8) and (S9).

2. With the obtained $p_{\text{end}}^d$ and $p_{\text{ave}}^d$, we define $u_d^{\ell+1}$ as

$$u_d^{\ell+1}(n_{i-d}, \cdots, n_i, \cdots, n_{i+d}) = u_d^\ell(n_{i-d}, \cdots, n_i, \cdots, n_{i+d}) - 2a \log \left[\frac{p_{\text{ave}}^d(\mathcal{F}_i[\mathcal{C}];i)}{p_{\text{end}}^d(\mathcal{F}_i[\mathcal{C}];i)} \frac{p_{\text{end}}^d(\mathcal{C};i)}{p_{\text{ave}}^d(\mathcal{C};i)}\right], \tag{S10}$$

as inferred from (S5). Here, $0 < a < 1$ is a parameter that one can tune in order to stabilize the convergence: When $a$ is close to 1 (resp. 0), we need less (resp. more) iterations in order to converge, but the calculation presents larger (resp. smaller) fluctuations. For the FA model presented in the main text, since the modified dynamics satisfies detailed balance, we replace $p_{\text{ave}}^d$ by $e^{u_d^\ell}(p_{\text{end}}^d)^2/p_{\text{eq}}^d$, according to (S6), where $u_d^\ell$ is the potential at the $\ell$th iteration and $p_{\text{eq}}^d$ is the equilibrium distribution truncated at range $d$. One then finds that setting

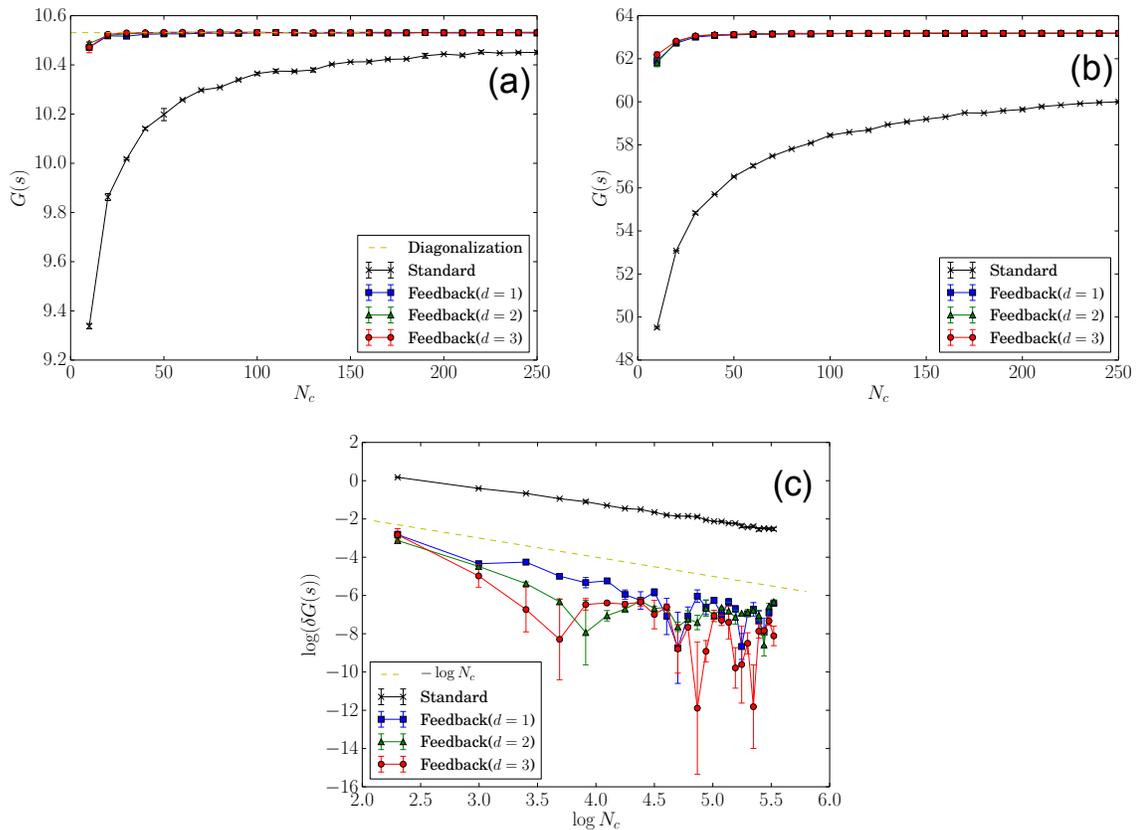

FIG. S1: The cumulant generating function $G(s)$ for $s = -1$ as a function of the number of copies $N_c$, obtained with the standard and with the feedback cloning algorithm (for $d = 1, 2, 3$). Here, $d$ is the range of the effective description (see (S7)). The system size is $L = 12$ **(a)** and $L = 72$ **(b)**, with a sufficiently large duration of the simulation ($\tau = 7500$). In the subfigure (a), we also plot the exact value obtained by numerical diagonalization as a yellow dashed line. For $L = 72$, although the standard method does not converge in the chosen window of $N_c$, the feedback method shows a very fast convergence in the large-$N_c$ limit, proving its advantage even for $d = 1$. **(c)** For $L = 12$, we plot the deviation $\delta G(s)$ of the estimator from the exact value (obtained by numerical diagonalization), in log-log scale. We also plot $-\log N_c$ as a straight yellow dashed line. We can see that the convergence of the estimator as $N_c$ increases is governed by the power law $\delta G(s) \sim N_c^{-1}$ (see Refs.[2, 3] for a systematic study). The prefactor of this power law is much smaller in the feedback algorithm than in the standard one.

$a = 1/2$ removes the first term of (S10) and we observed that this choice is the most stable numerically. The relation (S10) then writes:

$$u_d^{\ell+1}(n_{i-d}, \cdots, n_i, \cdots, n_{i+d}) = -\log \frac{p_{\text{end}}^d(\mathcal{F}_i[\mathcal{C}]; i)}{p_{\text{eq}}^d(\mathcal{F}_i[\mathcal{C}]; i)} \frac{p_{\text{eq}}^d(\mathcal{C}; i)}{p_{\text{end}}^d(\mathcal{C}; i)}. \tag{S11}$$

With this choice, the convergence of the CGF estimator is achieved at smaller $N_c$ than for the original algorithm. As seen on Fig. 1 in the main text, a few iterations of this procedure already provide an important improvement.

## II. COMPARISON OF THE STANDARD- AND THE FEEDBACK-POPULATION ALGORITHMS

In the main text, we show that the feedback algorithm improves the numerical estimation of the CGF around the dynamical phase transition. Here, we illustrate that it also greatly improves the results in the active phase, far from the transition point. In Fig. S1 we compare the CGF $G(s)$ obtained from both methods, as a function of $N_c$. The feedback is used with a single iteration, taking $a = 1$ in (S10). Even for nearest-neighbor interactions ($d = 1$), the results reach their large-$N_c$ limit much faster than with the original algorithm.

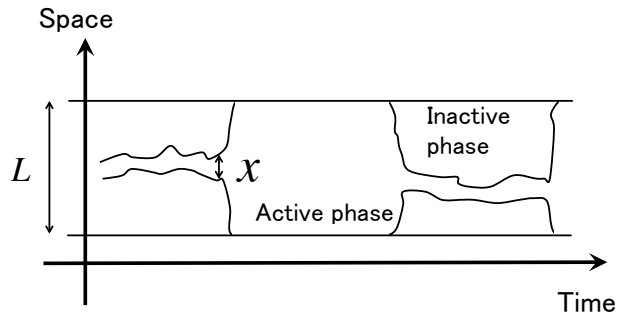

FIG. S2: The schematic of the interface model: $x(t)$ represents the width of the active region in the system at time $t$, for values of $s$ leading to phase coexistence.

## III. INTERFACE MODEL TO DERIVE THE ANALYTICAL RESULTS (9), (10) AND (11) OF THE MAIN TEXT

In this part, we analyze the interface model proposed in the main text. From the analysis, we derive the results (9), (10) and (11) of the main text, which describe the exponential scalings of the phase transition. In Sec. III A, we describe the model in its spatially discrete version, and present numerical evidences that the model possesses a dynamical phase transition. In Sec. III B, we take the continuum limit and derive a differential equation for the biased probability distribution. In Sec. III C, from the obtained differential equation, we derive the relations (9), (10) and (11) of the main text. In Sec. III D, we summarize the results obtained for our effective model.

### A. Interface model

#### 1. Model definition

As illustrated in Fig. S2, the effective model describes the dynamics of the width $x$ of the active domain. Such effective dynamics was proposed to describe the dynamics in inactive phase ($s > s_c^L$) in [4]. Here, we assume that this model can also describe the phase coexistence, i.e., the regime $s \sim s_c^L$ (see the main text) where the FA model presents a sole active domain. This will allow us to describe the finite-size scaling of the transition at a more detailed level than in [4]. We assume that the width $x$ is an integer that follows a random walk of rates

$$w(x \to x+1) = 2q, \qquad w(x \to x-1) = 2p, \tag{S12}$$

for $1 < x < L$. But different from [4], we take into account the finite size $L$ of the system through the following boundary conditions:

$$w(1 \to 0) = 0, \qquad w(1 \to 2) = 2q, \qquad w(L \to L+1) = 0, \qquad w(L \to L-1) = 2p, \tag{S13}$$

which ensure $x \geq 1$ and $x \leq L$. Note that $x \geq 1$ because a single active site cannot vanish, due to the kinetic constraint. We denote the trajectory of $x$ from $t=0$ to $t=\tau$ by $\omega$, and the trajectorial probability of $\omega$ by $P(\omega)$. For simplicity, we set $x(0) = 1$ but this does not affect the results in the large-time limit. As discussed in Section 4.1 of [4], the average rates for increasing and decreasing the domain size of active phase in FA model are given as $2c$ and $2c(1-c)$. We use these average rates to describe the interface dynamics here: We set $q = c$ and $p = c(1-c)$ with $0 < c < 1$, (implying the following inequalities: $0 < q < 1$, $0 < p < 1$ and $p < q$). The physical picture, that we repeat here for completeness, is the following: (i) a site from the empty region at the boundary can be filled with a rate $p$ equal to the creation rate $c$, while (ii) the occupied site at the boundary can be emptied with the annihilation rate $1-c$, times the probability $c$ that its neighbor in the active region is occupied. This reasoning assumes that the occupation measure is the same in the active region as in equilibrium, which is only approximate. The factors 2 come from the fact that the active region has two boundaries.

The time-averaged activity of the FA model corresponds to $\frac{1}{\tau}\bar{k}\int_0^\tau dt\, x(t)$ in the effective model, where $\bar{k} = 4c^2(1-c)$ is the mean density of the activity in FA per site and per unit time. The corresponding biased weight of trajectories for the effective model is thus:

$$P_s(\omega) = P(\omega) \exp\left(-s\bar{k}\int_0^\tau dt\, x(t)\right). \tag{S14}$$

We denote the corresponding biased expected value (with respect to $P_s(\omega)$) by $\langle \cdot \rangle_s$. We also define the corresponding cumulant generating function (or 'dynamical free energy') as

$$\Psi(s) = \lim_{\tau\to\infty} \frac{1}{\tau} \log \left\langle \exp\left(-s\bar{k}\int_0^\tau dt\, x(t)\right) \right\rangle_0. \tag{S15}$$

The derivative of $\Psi(s)$ gives (up to a sign) the expected value of the time-averaged activity:

$$\frac{\partial \Psi(s)}{\partial s} = -\bar{k} \lim_{\tau\to\infty} \left\langle \frac{1}{\tau} \int_0^\tau dt\, x(t) \right\rangle_s. \tag{S16}$$

We note that, within this interface model, the expected value of the activity is proportional to the average area of active phase per unit time $\langle \mathcal{A} \rangle_s \equiv \lim_{\tau\to\infty}(1/\tau)\langle \int_0^\tau dt\, x(t) \rangle_s$.

### 2. Numerical evidence for dynamical phase transitions in this interface model

For $s = 0$, the width of the interface $x(t)$ is equal to $L$ most of the time, so that the mean area is $\langle \mathcal{A} \rangle_{s=0} = \lim_{\tau\to\infty}(1/\tau)\langle \int_0^\tau dt\, x(t) \rangle_{s=0} \approx L$. On the other hand, in the $s \to \infty$ limit, $x(t)$ is close to 1 most of the time, leading to $\lim_{\tau\to\infty}(1/\tau)\langle \int_0^\tau dt\, x(t) \rangle_s \approx 1$. When $s$ is a finite value between these two extremal regimes, one will observe an intermediate behavior. We expect this behavior to be singular in the large-$L$ limit. This can be justified by an argument similar to the one used for the FA model [5]. We first note that $-\frac{1}{L}\Psi'(0) = \frac{1}{L}\bar{k}\langle \mathcal{A} \rangle_{s=0} = O(L^0) > 0$, while $\lim_{s\to\infty} \frac{1}{L}\Psi(s) = \frac{1}{L}\langle \mathcal{A} \rangle_{s\to\infty} = O(L^{-1})$. Then, because $\Psi'(s) = -\langle \mathcal{A} \rangle_s < 0$, we observe that $\frac{1}{L}\Psi(s)$ is a decreasing function of $s$. We thus have that $\forall s \geq 0, \frac{1}{L}\Psi(0) \geq \frac{1}{L}\Psi(s) \geq \lim_{s'\to\infty} \frac{1}{L}\Psi(s') = O(L^{-1})$. Taking the large-$L$ limit, we obtain that $\forall s \geq 0, \lim_{L\to\infty} \frac{1}{L}\Psi(s) = 0$. Since, as observed previously, $\lim_{L\to\infty} -\frac{1}{L}\Psi'(0) > 0$, this implies that $\lim_{L\to\infty} \frac{1}{L}\Psi(s)$ is non-analytic in $s = 0$.

To support this claim, we present on Fig. S3 numerical evidences that the interface model indeed possesses a dynamical phase transition. We set $c = 0.3$ and plot

$$X(\lambda) = \lim_{\tau\to\infty} \frac{1}{\tau} \left\langle \int dt\, \frac{x(t)}{L} \right\rangle_{s=\lambda/L} \tag{S17}$$

(as a function of $\lambda \equiv Ls$) for several values of $L$ on Fig. S3(a). We note that $X(\lambda)$ is proportional to $\partial\Psi/\partial\lambda$. We define a dynamical susceptibility as

$$\chi(\lambda) = -\frac{\partial X(\lambda)}{\partial \lambda}, \tag{S18}$$

(which is thus proportional to $\partial^2\Psi/\partial\lambda^2$). We also plot this quantity in Fig. S3(b). One observes the development of first-order dynamical phase transition as $L$ increases. To characterize it, we denote by $\kappa_{\text{eff}}$ the largest value of $\chi$ around the transition:

$$\kappa_{\text{eff}} \equiv \max_\lambda \chi(\lambda) \tag{S19}$$

and also we define $\lambda_{\max}$ as the value maximizing $\chi(\lambda)$:

$$\lambda_{\max} = \operatorname{argmax}_\lambda \chi(\lambda). \tag{S20}$$

If $\kappa_{\text{eff}}$ diverges as $\exp(\alpha L)$ for $L \to \infty$, then, we expect that the time scale ruling the dynamics close to the transition diverges also as $\exp(\alpha L)$. We evaluate numerically $\kappa_{\text{eff}}$ and we plot $\log \kappa_{\text{eff}}$ as a function of $L$ in Fig. S4. The results support an exponential behavior $\kappa_{\text{eff}} = O(e^{\alpha L})$ with a divergence rate $\alpha \approx 0.11$ for $c = 0.3$.

### B. Continuous space description

In this section, we take the continuous space limit of the random-walk effective model. As in Ref. [4], this leads to a second-order differential equation for the evolution of the biased probability, but here we take into account the boundary conditions (S13) that keep $x/L$ *finite*. This allows us to obtain analytical expressions describing the finite-size effects occurring in an exponentially small regime around the first-order transition, as discussed in the main text.

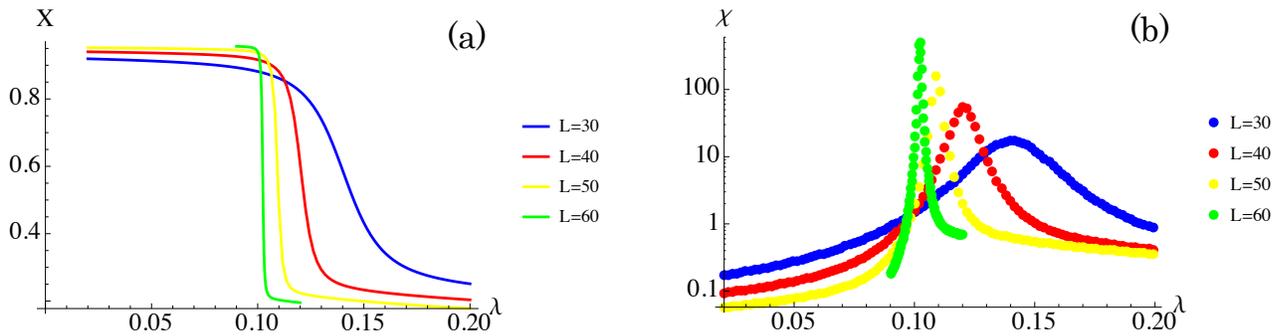

FIG. S3: **(a)** The average area $X(\lambda)$ defined in (S17) as a function of $\lambda = sL$. **(b)** The susceptibility $\chi(\lambda)$ defined in (S18) as a function of $\lambda$. Both functions are represented for several values of $L$, with $c = 0.3$. The results support the existence of a first-order transition as $L \to \infty$ at some finite critical $\lambda_c > 0$.

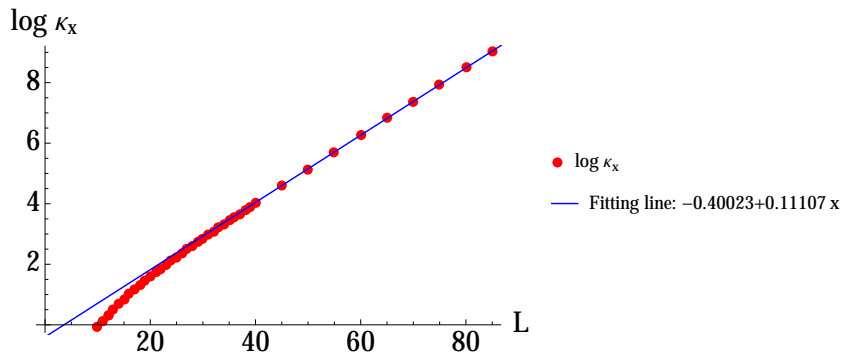

FIG. S4: Dependence in $L$ of $\log \kappa_{\text{eff}}$, with the susceptibility $\kappa_{\text{eff}}$ defined in (S19). The figure indicates that $\kappa_{\text{eff}} = O(e^{0.11L})$ for $c = 0.3$.

### 1. Bulk equation

We consider the eigenvalue problem for the dynamical free energy $\Psi(s)$, which is written as

$$2pP(x+1) + 2qP(x-1) - (2p + 2q + sx\overline{k})P(x) = \Psi(s)P(x), \tag{S21}$$

with the corresponding eigenvector $P(x)$. See for example Ref. [5] for a generic derivation of such eigenstate equations. For $s = 0$, the equation describes a probability-conserving asymmetric random walk, implying that the largest eigenvalue is 0, ensuring $\Psi(0) = 0$ as expected. The corresponding eigenvector for $s = 0$ is $P(x)|_{s=0} = (q/p)^x$. From this expression, we define

$$F = (1/2)\log(q/p). \tag{S22}$$

To symmetrize the walk, we then change variable through $Q(x) = (p/q)^{(x/2)}P(x)$ as in Ref. [4] and rewrite (S21) as

$$2\sqrt{pq}\left[Q(x+1) + Q(x-1) - 2Q(x)\right] - \left(2p + 2q - 4\sqrt{pq} + sx\overline{k}\right)Q(x) = \Psi(s)Q(x). \tag{S23}$$

We now focus on the large-$L$ regime and define a new variable $\tilde{x} = x/L$ with a new parameter $\lambda = sL$. We assume that $\tilde{Q}(\tilde{x}) \equiv LQ(\tilde{x}L)$ varies sufficiently smoothly with $\tilde{x}$ so that $L^2\left[\tilde{Q}(\tilde{x} + 1/L) + \tilde{Q}(\tilde{x} - 1/L) - 2\tilde{Q}(\tilde{x})\right]$ can be replaced by its second derivative with respect to $\tilde{x}$. We thus obtain the following second order differential equation

$$\frac{\partial^2 \tilde{Q}}{\partial \tilde{x}^2} - L^2 \mu^3 \tilde{x} \tilde{Q}(\tilde{x}) = L^2 \psi \tilde{Q}(\tilde{x}), \tag{S24}$$

with $\mu^3 = \lambda \overline{k}/(2\sqrt{pq})$ and

$$\psi = \frac{\Psi}{2\sqrt{pq}} + 2\left(\cosh F - 1\right). \tag{S25}$$

The general solution of (S24) is expressed as

$$\tilde{Q}(\tilde{x}) = a\,\mathrm{Ai}\left(L^2\left(\mu\tilde{x} + \frac{\psi}{\mu^2}\right)\right) + b\,\mathrm{Bi}\left(L^2\left(\mu\tilde{x} + \frac{\psi}{\mu^2}\right)\right), \tag{S26}$$

where Ai and Bi are the Airy functions. The parameters $a$ and $b$ are constants determined by the boundary conditions, that we discuss in the next subsection.

### 2. Boundary conditions

From the definition of the random-walk model, the boundary condition of $P(x)$ for $x = 1$ (i.e., $\tilde{x} = 1/L$) is written as

$$2pP(2) - 2qP(1) - s\bar{k}P(1) = \Psi P(1). \tag{S27}$$

We rewrite this relation as

$$\frac{P(2) - P(1)}{P(2) + P(1)} = -\frac{2p - \Psi - sK - 2q}{2p + \Psi + sK + 2q}. \tag{S28}$$

Using the relation $P(\tilde{x}L) = e^{\tilde{x}LF}\tilde{Q}(\tilde{x})$, we obtain (for large $L$) the boundary condition of $\tilde{Q}$ for $\tilde{x} = 0$ as

$$\frac{\tilde{Q}'(0)}{\tilde{Q}(0)} = LB_0(q,p,\psi) \tag{S29}$$

with

$$B_0(q,p,\psi) = -2\frac{2p - \Psi(\psi) - 2q}{2p + \Psi(\psi) + 2q} - F. \tag{S30}$$

In (S29) and hereafter we use the notation $\cdot'$ to denote the derivative with respect to $\tilde{x}$.

Next, we consider the boundary condition of $P$ for $x = L$ (i.e., $\tilde{x} = 1$), which is written as

$$2qP(L-1) - 2pP(L) - \lambda\bar{k}P(L) = \Psi P(L). \tag{S31}$$

We rewrite it as

$$\frac{P(L) - P(L-1)}{P(L) + P(L-1)} = \frac{-2p + 2q - \Psi - \lambda\bar{k}}{2p + 2q + \Psi + \lambda\bar{k}}, \tag{S32}$$

which leads to the boundary condition of $\tilde{Q}(\tilde{x})$ for $\tilde{x} = 1$

$$\frac{\tilde{Q}'(1)}{\tilde{Q}(1)} = LB_1(q,p,\psi,\mu) \tag{S33}$$

with

$$B_1(q,p,\psi,\mu) = 2\frac{-2p + 2q - \Psi(\psi) - 2\sqrt{qp}\mu^3}{2p + 2q + \Psi(\psi) + 2\sqrt{qp}\mu^3} - F(q,p). \tag{S34}$$

### 3. Summary: problem to solve

We have to determine the integration constants $a$, $b$ and the eigenvalue $\psi$ in (S26) from the boundary conditions (S29) and (S33). In this determination, we regard $\Psi$ as a function of $\psi$, as read from (S25). We set $b = 1$ without loss of generality because the normalization of $\tilde{Q}(\tilde{x})$ does not matter. In general, one finds many solutions for $a$ and $\psi$. We focus on the pair maximizing $\psi$.

**C. Derivation of the expressions describing the finite-size scaling**

From the analysis explained below, we derive the relations (9), (10) and (11) of the main text. Since the analysis is technical, we summarize the results in Sec. III D.

*1. Exponential divergence in the boundary condition*

We substitute the general solution (S26) into the boundary conditions (S29) and (S33), and solve the resulting equations with respect to $a$. We denote by $a_0$ and $a_1$ the solutions obtained from the boundary condition of $\tilde{x} = 0$ and $\tilde{x} = 1$ respectively. These are

$$a_0 = \frac{L^{2/3}\mu \operatorname{Bi}'\left(L^{2/3}\psi/\mu^2\right) - L \operatorname{Bi}\left(L^{2/3}\psi/\mu^2\right) B_0(p,q,\psi)}{-L^{2/3}\mu \operatorname{Ai}'\left(L^{2/3}\psi/\mu^2\right) + L \operatorname{Ai}\left(L^{2/3}\psi/\mu^2\right) B_0(p,q,\psi)} \tag{S35}$$

and

$$a_1 = \frac{L^{2/3}\mu \operatorname{Bi}'\left(L^{2/3}\left(\psi/\mu^2 + \mu\right)\right) - L \operatorname{Bi}\left(L^{2/3}\left(\psi/\mu^2 + \mu\right)\right) B_1(p,q,\psi,\mu)}{-L^{2/3}\mu \operatorname{Ai}'\left(L^{2/3}\left(\psi/\mu^2 + \mu\right)\right) + L \operatorname{Ai}\left(L^{2/3}\left(\psi/\mu^2 + \mu\right)\right) B_1(p,q,\psi,\mu)}. \tag{S36}$$

The desired $\psi$ is then determined from the condition

$$a_0 = a_1. \tag{S37}$$

We look for the solution satisfying $-\mu^3 < \psi < 0$ (and one can check that this assumption is valid). This implies that $a_1$ can diverge exponentially in the $L \to \infty$ limit, as explained below. We use the following asymptotic behaviors of the Airy functions at large $y > 0$:

$$\operatorname{Ai}(y) = \exp\left(-\frac{2}{3}y^{3/2}\right)\left[\frac{1}{2\sqrt{\pi}}y^{-1/4} + O(y^{-7/4})\right], \tag{S38}$$

$$\operatorname{Ai}'(y) = -\exp\left(-\frac{2}{3}y^{3/2}\right)\left[\frac{1}{2\sqrt{\pi}}y^{1/4} + O(y^{-5/4})\right], \tag{S39}$$

$$\operatorname{Bi}(y) = \exp\left(\frac{2}{3}y^{3/2}\right)\left[\frac{1}{\sqrt{\pi}}y^{-1/4} + O(y^{-7/4})\right], \tag{S40}$$

$$\operatorname{Bi}'(y) = \exp\left(\frac{2}{3}y^{3/2}\right)\left[\frac{1}{\sqrt{\pi}}y^{1/4} + O(y^{-5/4})\right]. \tag{S41}$$

These behaviors indicate that $a_1$ diverges as $\exp\left[\frac{4}{3}L(\psi/\mu^2 + \mu)^{3/2}\right]$, unless there is a compensation between the polynomial prefactors of this exponential. We also remark that there are no exponential divergences in the Airy functions appearing in the expression of $a_0$. Thus, $a_0$ is finite unless its denominator vanishes. For general values of the parameters $\mu$ and $\psi$, we thus find that $a_1$ diverges exponentially as $L \to \infty$ while $a_0$ does not. For $a_0$ to be equal to $a_1$, there are thus two possibilities: $\mu$ and $\psi$ have to be fine tuned so that either (i) the denominator of $a_0$ is close to 0, or (ii) the numerator of $a_1$ is close to 0. Here, "close to" means that the deviation from 0 is exponentially small in $L$ with the correct rate, that we determine below.

*2. $\psi$ canceling the exponential divergence*

To discuss these two possibilities more quantitatively, we define $\psi_0(\mu)$ as the value of $\psi$ making the denominator of $a_0$ to be exactly 0:

$$-L^{2/3}\mu \operatorname{Ai}'\left(L^{2/3}\psi_0(\mu)/\mu^2\right) + L \operatorname{Ai}\left(L^{2/3}\psi_0(\mu)/\mu^2\right) B_0(p,q,\psi_0(\mu)) = 0, \tag{S42}$$

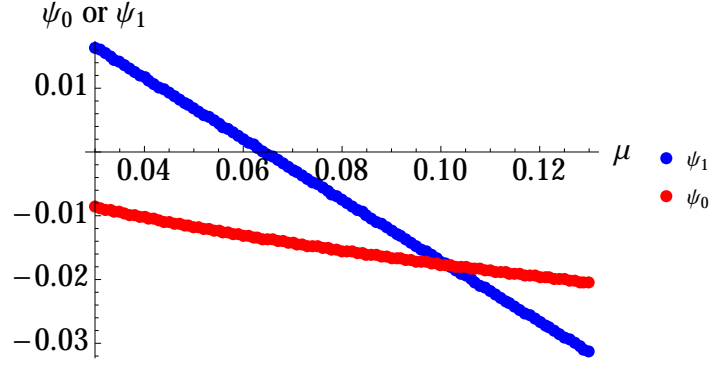

FIG. S5: Numerical examples of the functions $\psi_0(\mu)$ and $\psi_1(\mu)$ as a function of $\mu$ for $L = 50$, $q = c$ and $p = c(1-c)$ with $c = 0.3$. These values are determined by solving numerically the equations (S42) and (S43).

and $\psi_1(\mu)$ as the value of $\psi$ making the numerator of $a_1$ to be exactly 0:

$$L^{2/3}\mu\mathrm{Bi}'\left(L^{2/3}\left(\psi_1(\mu)/\mu^2+\mu\right)\right) - L\mathrm{Bi}\left(L^{2/3}\left(\psi_1(\mu)/\mu^2+\mu\right)\right)B_1(p,q,\psi_1(\mu),\mu) = 0. \tag{S43}$$

For finite $L$, the actual value of $\psi$ will be close to either $\psi_0$ or $\psi_1$, as we discuss below. We plot the functions $\psi_0(\mu)$ and $\psi_1(\mu)$ in Fig. S5 as a function of $\mu$ for a numerical example. These two functions are different in general.

The function $\psi(\mu)$ that makes $a_0 = a_1$ is determined as follows: For a given $\mu$, we first compare $\psi_1$ and $\psi_0$ and choose the largest value. For example, when $\psi_1 > \psi_0$, we choose $\psi_1$, and $\psi(\mu)$ is given as

$$\psi(\mu) = \psi_1(\mu) + O(e^{-\frac{4}{3}L(\psi_1/\mu^2+\mu)^{3/2}}). \tag{S44}$$

and when $\psi_0 > \psi_1$, we choose $\psi_0$, and $\psi(\mu)$ is given as

$$\psi(\mu) = \psi_0(\mu) + O(e^{-\frac{4}{3}L(\psi_0/\mu^2+\mu)^{3/2}}), \tag{S45}$$

where the order of the deviations is inferred from the cancellation of the exponential divergence of $a_1$.

### 3. A phase transition

At the point where $\psi_1$ and $\psi_0$ take the same value, $\psi$ changes from $\psi_1$ to $\psi_0$ as $\mu$ increases (see Fig. S5). Hence, the derivative of $\psi$ shows a discontinuity at the crossing point in the $L \to \infty$ limit, which can be clearly seen, for example, in Fig. S5. We thus define the special value $\mu^*$ from the condition

$$\psi_1(\mu^*) = \psi_0(\mu^*) \equiv \psi^*. \tag{S46}$$

We stress that $\mu^*$ depends on $L$. We denote by $\lambda^*$ the value of $\lambda$ corresponding to $\mu^*$ via

$$\lambda = (2\sqrt{pq}/\overline{k})\mu^3. \tag{S47}$$

It will turn out that this definition of $\lambda^*$ is equivalent to the one maximizing the second derivative of free energy $\Psi$, or $\psi$ (see for instance Eq. (S20)).

### 4. Finite-size scaling

We now study quantitatively the finite-size scaling around this phase transition. Consider a small deviation of $\mu$ from $\mu^*$:

$$\mu = \mu^* + \delta\mu. \tag{S48}$$

We denote the corresponding deviation of $\psi$ from $\psi^*$ as

$$\psi = \psi^* + \delta\psi. \tag{S49}$$

We rewrite the equation (S37) as $1 = a_1/a_0$, and expand the right-hand side with respect to $\delta\mu$ and $\delta\psi$. From the definition of $\psi_0$ and $\psi_1$, we know that $a_1|_{\delta\mu=0,\delta\psi=0} = 0$ and $1/a_0|_{\delta\mu=0,\delta\psi=0} = 0$. Thus, the expansion is written as

$$1 = \frac{a_1}{a_0} = \frac{1}{2}\left[A_{\mu,\mu}\delta\mu^2 + A_{\psi,\psi}\delta\psi^2 + (A_{\mu,\psi} + A_{\psi,\mu})\delta\mu\delta\psi\right] + O(\delta\mu^3) + O(\delta\psi^3) + O(\delta\mu^2\delta\psi) + O(\delta\mu\delta\psi^2). \tag{S50}$$

where we defined

$$\left.\frac{\partial a_1}{\partial\mu}\frac{\partial a_0^{-1}}{\partial\mu}\right|_{\mu=\mu^*,\psi=\psi^*} = A_{\mu,\mu}, \qquad \left.\frac{\partial a_1}{\partial\psi}\frac{\partial a_0^{-1}}{\partial\psi}\right|_{\mu=\mu^*,\psi=\psi^*} = A_{\psi,\psi}, \tag{S51}$$

$$\left.\frac{\partial a_1}{\partial\mu}\frac{\partial a_0^{-1}}{\partial\psi}\right|_{\mu=\mu^*,\psi=\psi^*} = A_{\mu,\psi}, \qquad \left.\frac{\partial a_1}{\partial\psi}\frac{\partial a_0^{-1}}{\partial\mu}\right|_{\mu=\mu^*,\psi=\psi^*} = A_{\psi,\mu}. \tag{S52}$$

We note that all these coefficients ($A_{\mu,\mu}$, $A_{\mu,\psi}$, $A_{\psi,\mu}$, $A_{\psi,\psi}$) are of order $O(e^{\frac{4}{3}L(\psi^*/(\mu^*)^2+\mu^*)^{3/2}})$. In order to solve Eq. (S50), we thus set

$$\delta\tilde{\mu} = \sqrt{\frac{A_{\mu,\mu}}{2}}\delta\mu, \qquad \delta\tilde{\psi} = \sqrt{\frac{A_{\psi,\psi}}{2}}\delta\psi. \tag{S53}$$

Due to the fast divergence of $A_{\mu,\mu}$ and $A_{\psi,\psi}$, expanding (S50) in powers of $\delta\tilde{\mu}$ and $\delta\tilde{\psi}$, the terms in $O(\delta\mu^3)$, $O(\delta\psi^3)$, $O(\delta\psi^2\delta\mu)$ and $O(\delta\psi\delta\mu^2)$ can be neglected. Thus, (S50) is written as

$$1 = \delta\tilde{\mu}^2 + \delta\tilde{\psi}^2 + \frac{(A_{\mu,\psi} + A_{\psi,\mu})}{\sqrt{A_{\psi,\psi}A_{\mu,\mu}}}\delta\tilde{\mu}\delta\tilde{\psi}. \tag{S54}$$

Denoting $C = \frac{(A_{\mu,\psi}+A_{\psi,\mu})}{\sqrt{A_{\psi,\psi}A_{\mu,\mu}}}$, this equation is solved as

$$\delta\tilde{\psi} = \frac{1}{2}\left[-C\delta\tilde{\mu} \pm \sqrt{(C^2-4)\delta\tilde{\mu}^2 + 4}\right], \tag{S55}$$

The derivative $\frac{\partial(\delta\tilde{\psi})}{\partial(\delta\tilde{\mu})}$ is

$$\frac{\partial(\delta\tilde{\psi})}{\partial(\delta\tilde{\mu})} = \frac{1}{2}\left[-C \pm \frac{(C^2-4)\delta\tilde{\mu}}{\sqrt{(C^2-4)\delta\tilde{\mu}^2+4}}\right]. \tag{S56}$$

The undetermined sign $\pm$ in this expression is obtained from the sign of $(C^2 - 4)$: To see this, we consider the derivative of $\frac{\partial(\delta\tilde{\psi})}{\partial(\delta\tilde{\mu})}$ at $\delta\tilde{\mu} = 0$, which reads

$$\left.\frac{\partial^2(\delta\tilde{\psi})}{\partial(\delta\tilde{\mu})^2}\right|_{\delta\tilde{\mu}=0} = \pm\frac{C^2-4}{2}. \tag{S57}$$

As justified below, this term is proportional to $\partial^2\Psi/\partial\lambda^2$, which should be positive by convexity of the CGF. This fixes the undetermined sign in (S55) by replacing $\pm(C^2-4)$ by $|(C^2-4)|$.

### 5. Connection to the original finite-size scaling

We now write down the result of our previous analysis for $\Psi$ as function of $\lambda$. From (S47), setting $\lambda = b\mu^3$ with $b \equiv 2\sqrt{pq}/\bar{k}$, one has

$$\frac{\partial\Psi}{\partial\lambda} = \frac{1}{3b\mu^2}\frac{\partial\Psi}{\partial\mu} = \frac{\sqrt{qp}}{3b\mu^2}\frac{\partial\psi}{\partial\mu} = \frac{\sqrt{qp}}{3b\mu^2}\frac{\partial\delta\psi}{\partial\delta\mu} = \sqrt{\frac{A_{\mu,\mu}}{A_{\psi,\psi}}}\frac{\sqrt{qp}}{3b\mu^2}\frac{\partial\delta\tilde{\psi}}{\partial\delta\tilde{\mu}}. \tag{S58}$$

(This justifies the connection between the signs of $\partial^2 \Psi / \partial \lambda^2$ and $\frac{\partial^2 (\delta \tilde{\psi})}{\partial (\delta \tilde{\mu})^2}$ used above.) We thus obtain

$$\frac{\partial \Psi}{\partial \lambda} = \sqrt{qp \frac{A_{\mu,\mu}}{A_{\psi,\psi}}} \frac{1}{6b(\mu^*)^2} \left[ -C + \frac{|C^2 - 4|\delta \tilde{\mu}}{\sqrt{(C^2 - 4)\delta \tilde{\mu}^2 + 4}} \right]. \tag{S59}$$

Furthermore, $\delta \tilde{\mu}$ and $\delta \lambda = \lambda - \lambda^*$ are related by

$$\delta \lambda = 3b(\mu^*)^2 \sqrt{\frac{2}{A_{\mu,\mu}}} \delta \tilde{\mu}. \tag{S60}$$

From this, the maximum of $\partial^2 \Psi / \partial \lambda^2$ around the transition point is

$$\kappa = \sqrt{qp \frac{A_{\mu,\mu}^2}{2 A_{\psi,\psi}}} \frac{1}{36 b^2 (\mu^*)^4} |C^2 - 4|. \tag{S61}$$

Besides, the derivative $\partial \Psi / \partial \lambda$ of the CGF is

$$\frac{\partial \Psi}{\partial \lambda} = -A + \frac{\kappa \delta \lambda}{\sqrt{B(\kappa \delta \lambda)^2 + 1}}, \tag{S62}$$

with

$$A = C \sqrt{qp \frac{A_{\mu,\mu}}{A_{\psi,\psi}}} \frac{1}{6b(\mu^*)^2}, \qquad B = \frac{36 b^2 (\mu^*)^4}{C^2 - 4} \frac{A_{\psi,\psi}}{qp A_{\mu,\mu}}. \tag{S63}$$

These results constitute the relation (11) of the main text. From this expression, one can confirm that $\mu^*$, defined from (S46), gives (through (S47)) the point $\lambda^*$ which maximizes $\partial^2 \Psi(\lambda) / \partial \lambda^2$. Furthermore, the scaling coefficient $\kappa$ is of order $O(e^{2L/3(\psi^*/(\mu^*)^2 + \mu^*)^{3/2}})$, leading to

$$\frac{\partial^2 \Psi}{\partial \lambda^2} \sim e^{2L/3(\psi^*/(\mu^*)^2 + \mu^*)^{3/2}}. \tag{S64}$$

### 6. Simplifications in $L \to \infty$ limit

We finally determine the $L \to \infty$ limit of the exponential divergence rate $2/3 \left[ \psi^*/(\mu^*)^2 + \mu^* \right]^{3/2}$ of the susceptibility in (S64). First, we notice that the first and second terms in the equation (S42) for $\psi_0$ are of different order: the first term is of order $L^{2/3}$ and the second one is of order $L$. This indicates that

$$L^{2/3} \frac{\psi_0(\mu)}{\mu^2} \to \alpha_0, \tag{S65}$$

where $\alpha_0$ is the first positive zero of the Airy function Ai. This leads to $\lim_{L \to \infty} \psi_0(\mu) = 0$ and $\lim_{L \to \infty} \psi^* = 0$ from the definition of $\psi^* = \psi_0(\mu^*) = \psi_1(\mu^*)$. In order to calculate $\lim_{L \to \infty} \mu^*$, we use the condition $\psi_1(\mu^*) = 0$ in (S43). Using the asymptotic behaviors (S38-S41) of the Airy functions, we obtain

$$(\mu_\infty^*)^3 - (\mu_\infty^*)^{3/2} B_1(q, p, 0, \mu_\infty^*) = 0, \tag{S66}$$

which is the relation (9) of the main text (where $\mu$ denotes $\mu_\infty^*$). Now, using $\lim_{L \to \infty} \psi_0(\mu) = 0$ in the expression $2/3 (\psi^*/(\mu^*)^2 + \mu^*)^{3/2}$, we finally obtain

$$\alpha = \frac{2}{3} (\mu_\infty^*)^{3/2}. \tag{S67}$$

This is the relation (10) of the main text.

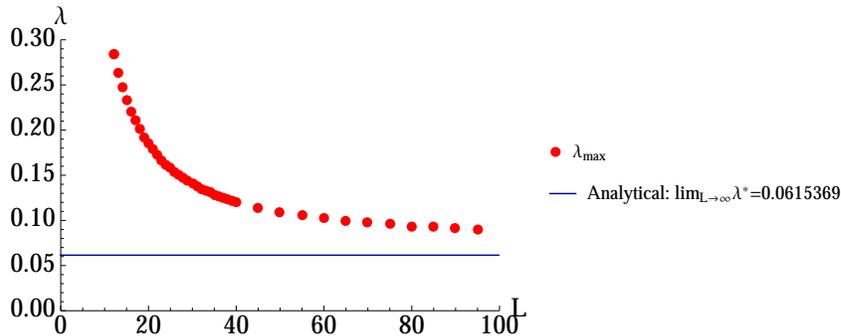

FIG. S6: The critical point $\lambda_\infty^*$ (blue solid line) and the corresponding finite-size $\lambda_{\max}$ (red dots), defined in (S20), as a function of $L$.

### D. Summary

#### 1. Analytical results

We defined the special value $\lambda^*$ of $\lambda$ through (S47), where the corresponding $\mu^*$ is defined in (S46) for large but finite $L$. This value indeed maximizes $\partial^2 \Psi/\partial \lambda^2$ around the transition point, and the derivative of $\Psi$ with respect to $\lambda$ becomes discontinuous at $\lambda^*$ in the $L \to \infty$ limit:

$$\lim_{L\to\infty} \frac{\partial \Psi}{\partial \lambda}\bigg|_{\lambda \downarrow \lambda^*} \neq \lim_{L\to\infty} \frac{\partial \Psi}{\partial \lambda}\bigg|_{\lambda \uparrow \lambda^*}. \tag{S68}$$

The second derivative of $\Psi$ at $\lambda^*$ then diverges exponentially as $L \to \infty$:

$$\frac{\partial^2 \Psi}{\partial \lambda^2}\bigg|_{\lambda = \lambda^*} \sim \exp(\alpha L), \tag{S69}$$

where the divergence rate $\alpha$ is given as (S67). In this expression, $\mu_\infty^* \equiv \lim_{L \to \infty} \mu^*$ is determined from (S66). The expressions (S66) and (S67) correspond to (9) and (10) in the main text.

Furthermore, defining $\delta\lambda$ (as $\delta\lambda = \lambda - \lambda^*$), $\kappa$ (by (S61)), and $A$, $B$ (by (S63)), we obtain the expression of $\partial\Psi/\partial\lambda$ around the transition point as (S62), which is nothing but (11) in the main text. We note that the parameters $A_{\mu,\mu}$, $A_{\psi,\psi}$ in the expressions (S61) and (S63) are of order $O(e^{\frac{4}{3}L(\psi^*/(\mu^*)^2+\mu^*)^{3/2}})$, which ensures the exponential divergence of $\kappa$ and the absence of such divergence in the constants $A$ and $B$.

#### 2. Numerical example

Setting $q = c$ and $p = c(1-c)$ with $c = 0.3$, we obtain $\mu_\infty^* \approx 0.31377$ and $\alpha \approx 0.117173$ by solving the previous equations numerically. The corresponding value of $\lambda^*$ in the $L \to \infty$ limit, that we denote $\lambda_\infty^*$, is $\lambda_\infty^* \approx 0.0615369$. To test these analytical results, we refer to the numerical results of Fig. S4 that are obtained before taking the continuum limit by direct diagonalization of the evolution operator. Identifying the rate of the exponential divergence from the numerical data up to system size $L = 90$, one finds $\alpha \approx 0.11107$, which is very close to the analytical value given above. Furthermore, on Fig. S6, we plot $\lambda_{\max}(L)$ defined from (S20) as a function of $L$. The results are compatible with $\lim_{L \to \infty} \lambda_{\max}(L) = \lambda_\infty^*$.

## IV. EFFECTIVE INTERACTIONS AROUND THE DYNAMICAL PHASE TRANSITION POINTS

Here, we study numerically the effective interactions around the dynamical phase transition. For the effective potential $U$, we consider four interactions: one-body, nearest-neighbor, second nearest-neighbor, and three-body. In order to calculate the strength of these interactions, we consider a configuration where all spins are 0 except for the

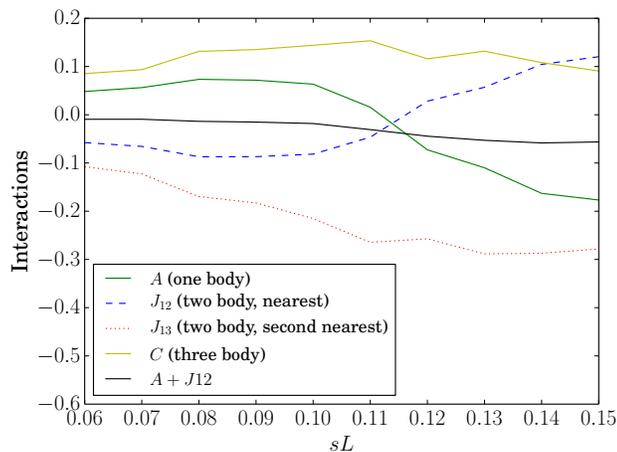

FIG. S7: Interaction coefficients of $U$ defined as (S70) around the dynamical phase transition. We obtain these coefficients from the results of feedback algorithm with $N_c = 400$ and $\tau = 15000$.

three adjacent ones on the site $i$, $i \pm 1$. For this configuration, $U$ is written as

$$U(0, \cdots, n_{i-1}, n_i, n_{i+1}, \cdots, 0) = A\left(n_{i-1} + n_i + n_{i+1}\right) + J_{12}\left(n_{i-1}n_i + n_i n_{i+1}\right) + J_{13} n_{i-1} n_{i+1} + C n_{i-1} n_i n_{i+1} + \text{const.,} \tag{S70}$$

where $A$, $J_{12}$, $J_{13}$ and $C$ are the coefficients representing the strength of the four interactions. Using the effective potential obtained from the implementation of the algorithm for $d = 4$, we estimate these parameters $A$, $J_{12}$, $J_{13}$ and $C$, and we plot them as a function of $sL$ in Fig S7. One observes that the nearest-neighbor interactions $J_{12}$ and the one-body potential $A$ change their signs around the transition point, but the sum of these is always close to 0. The barrier to increase or decrease the size of the active region is thus small throughout the finite-size rounding. Furthermore, especially close to the transition point, the short-range interactions are not important, since they are close to 0. But we note that, even if the long-range nature of the effective interactions around the phase coexistence is important, still, taking effective interactions with $d = 4$ did improve significantly the efficiency of the algorithm (see Fig. 1 of the main text).



\end{document}